\documentclass[9pt,preprint,natbib209]{aastex}
\usepackage{wasysym}
\begin{document}

\shorttitle{LMXBs with SSs } \shortauthors{Zhu et al.}

\title{Low-mass X-ray Binaries with Strange Quark Stars}
\author{Chunhua Zhu\altaffilmark{1,2,3}, Guoliang L\"{u}\altaffilmark{1,2},
 Zhaojun Wang\altaffilmark{1,2}, Jinzhong Liu\altaffilmark{3}}
\email{$^\dagger$chunhuazhu@sina.cn}

\altaffiltext{1}{School of Physical Science and Technology, Xinjiang
University, Urumqi, 830046, China.}

\altaffiltext{2}{Xinjiang University-National Astronomical
Observatories Joint Center for Astrophysics, Urumqi, 830046, China}

\altaffiltext{3}{National Astronomical Observatories / Xinjiang
Observatory, the Chinese Academy of Sciences, Urumqi, 830011, China}

\begin{abstract}
Strange quark stars (SSs) may originate from accreting neutron stars
(NSs) in low-mass X-ray binaries (LMXBs). Assuming that conversion
of NS matter to SSs occurs when the core density of accreting NS
reaches to the density of quark deconfinement, $\sim 5 \rho_0$,
where $\rho_0\sim 2.7\times 10^{14}$g cm$^{-3}$ is nuclear
saturation density, we investigate LMXBs with SSs (qLMXBs). In our
simulations, about 1\permil --- 10\% of LMXBs can produce SSs, which
greatly depends on the masses of nascent NSs and the fraction of
transferred matter accreted by the NSs. If the conversion does not
affect binaries systems, LMXBs evolve into qLMXBs. We find that some
observational properties (spin periods, X-ray luminosities and
orbital periods) of qLMXBs are similar with those of LMXBs, and it
is difficult to differ them. If the conversion disturbs the binaries
systems, LMXBs can produce isolated SSs.  These isolated SSs could
be submillisecond pulsars, and their birthrate in the Galaxy is
$\sim$5--70 per Myr.

\end{abstract}

\section{Introduction}
\label{sec:introduction} There are at least three different kinds of
compact stars in the Universe: white dwarfs (WDs), neutron stars
(NSs), and black holes. \cite{Witten1984} suggested a possible
existence of compact objects consisting of strange quark matter. Due
to strange quark matter being absolutely stable, \cite{Haensel1986}
and \cite{Alcock1986} pointed out that NSs almost would be made of
strange matter and not neutrons. However, \cite{Alpar1987}
considered that glitching radio-pulsars are NSs and not strange
quark stars (SSs). \cite{Madsen1988} suggested that SSs can not be
formed directly in supernovae\footnote{\cite{Dai1995} and
\cite{Xu2001} suggested that SSs can be formed directly during or
shortly after some supernovae explosion when the central density of
the proto-NSs is high enough.}, or less they would eventually
contaminate the entire Galaxy. \cite{Kluzniak1994} suggested that
SSs could exist as millisecond pulsars. Due to the fast rotation and
thermonuclear bursts, \cite{Li1999} suggested that the SAX
J1808.4-3658 is a good SS candidate. These SSs can be formed in
low-mass X-ray binaries (LMXBs) via an accretion-triggered phase
transition of NS matter to SS matter \citep{Cheng1996}.

The phase transition requires the formation of a strange matter seed
in the NS. The strange matter is produced through the neutron matter
at a critical density. \cite{Serot1987} pointed out that the central
density of an 1.4 $M_\odot$ NS with a rather stiff equation of state
is sufficiently lower than the critical density. Based on the modern
equations of state in \cite{Wiringa1988}, \cite{Cheng1996} estimated
that the NSs with 1.4 $M_\odot$ must accrete matter of $\sim 0.5
M_\odot$ in order that their central densities reach the
deconfinement density. Once the above condition is satisfied, the
phase transition occurs.

\cite{Olinto1987} proposed that the process of the strange matter
swallowing the neutron matter is a slow mode. However,
\cite{Horvath1988} showed that it is hydrodynamically unstable.
\cite{Cheng1996} proposed that the conversion of neutron matter
should proceed in a detonation mode and could be accompanied by a
gamma-ray burst. \cite{Ouyed2002} suggested that there is a
quark-nova when the core of a NS (having experienced a transition to
an up and down quark phase) shrinks into the equilibrated quark
object after reaching strange quark matter saturation density (where
a composition of up, down and strange quarks is the favored state of
matter). In their model, the energy released as radiation in a
quark-nova is up to $10^{53}$ ergs. \cite{Ouyed2011} proposed that
the quark novae in LMXBs may be the engines of short gamma-ray
bursts.

Based on the above descriptions, it is possible that SSs originate
from the hydrodynamically unstable conversion or the slow conversion
in LMXBs. Using standard equation of states of neutron-rich matter,
\cite{staff2006} considered that the density of quark deconfinement
is $\sim 5 \rho_0$, where $\rho_0\sim 2.7\times 10^{14}$g cm$^{-3}$
is nuclear saturation density. According to the equation of states
in \cite{Akmal1998}, the gravitational mass of a NS is $\sim 1.8
M_\odot$ in order to reach $5\rho_0$. Therefore, it is very
important for our understanding of SSs' formation to study the mass
evolution of NSs in LMXBs.

In this work, by simulating the interaction of a magnetized NS with
its environment and utilizing a population synthesis code, we focus
on the mass change of NSs in LMXBs and the possibility from NSs
converting SSs in LMXBs, and investigate the properties of LMXBs
with SSs (qLMXBs). In Section 2, we present our assumptions and
describe some details of the modelling algorithm. In Section 3, we
discuss the main results and the effects of different parameters. In
Section 4, the main conclusions are given.

\section{Model}
For the simulation of binary evolution, we use rapid binary star
evolution code BSE \citep{Hurley2002} which was updated by
\cite{Kiel2006}. In interacting binaries, NSs can be formed via
three channels\citep[e.g.,][]{Ivanova2008,Kiel2008}: (i)
Core-collapse supernovae (CCSN) for a star; (ii)Evolution induced
collapse (EIC) of a helium star with a mass between $1.4$ and $2.5
M_{\odot}$ in which the collapse is triggered by electron capture on
$^{20}$Ne and $^{24}$Mg \citep{Miyaji1980}; (iii) Accretion-induced
collapses (AIC) for an accreting ONeMg WD whose mass reaches the
Chandrasekhar limit. Response of accreting ONeMg WD is treated in
the same way as the evolution of CO WD \citep[see details
in][]{Lu2009}.
\subsection{Mass of Nascent NS}
\begin{figure}
\includegraphics[totalheight=3.3in,width=3.3in,angle=-90]{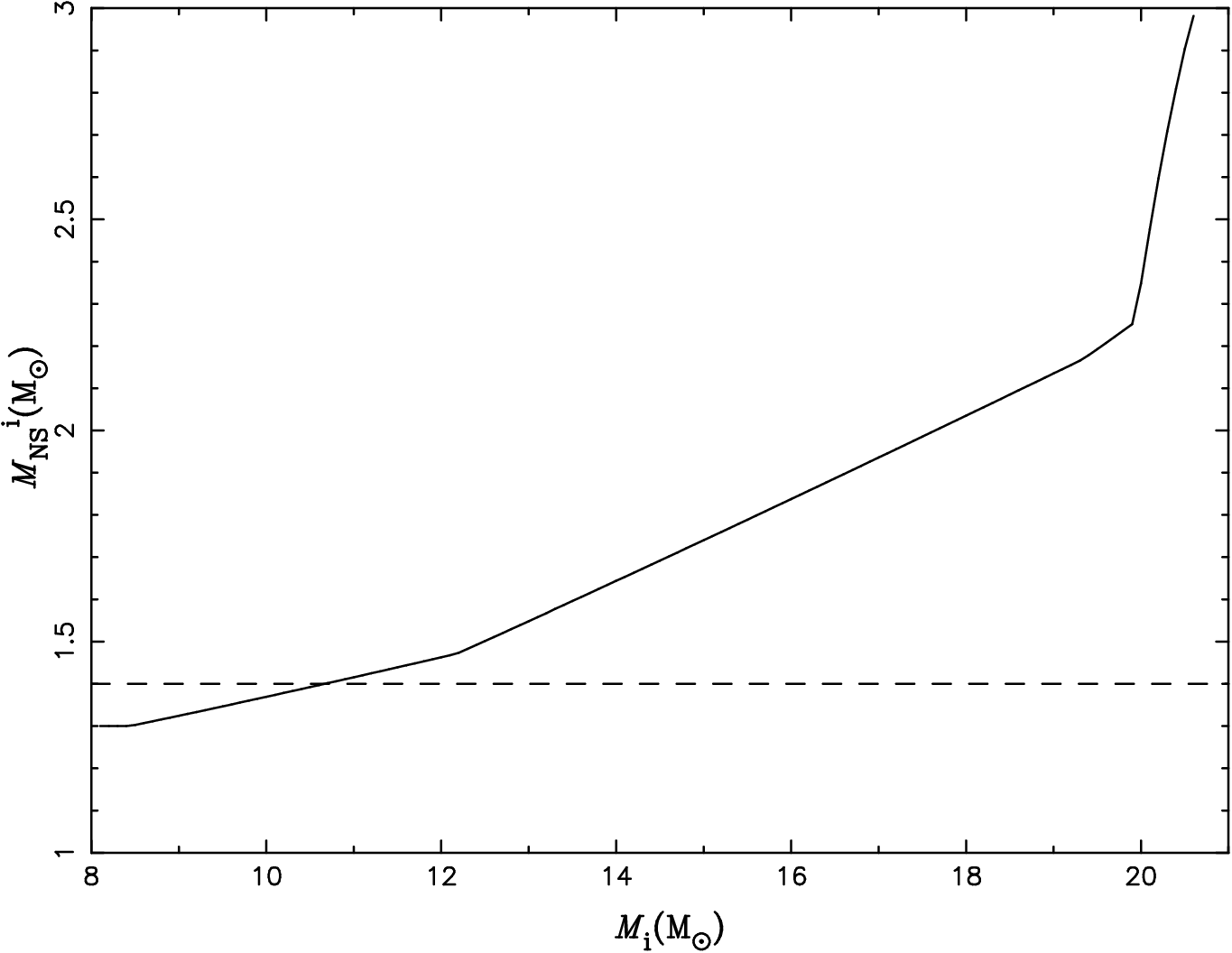}
\caption{Masses of nascent NSs via CCSN vs. stellar initial masses.
The solid line comes from \cite{Hurley2000}, and the dashed line
means that masses of nascent NSs are equal to $1.4 M_\odot$.
}\label{fig:nsm}
\end{figure}

Possibly the mass is one of the most important properties of NSs.
However, the mass distribution of nascent NSs is not yet well known.
In BSE code, the gravitational mass of a nascent NS via CCSN depends
on the mass of the CO-core at the time of supernova
\citep{Hurley2000}. Figure \ref{fig:nsm} shows the masses of nascent
NSs forming from different initial masses. Some authors assumed that
the initial masses of NSs ($M_{\rm NS}^{\rm i}$) are 1.4 $M_\odot$
in their works
\citep[e.g.,]{Ergma1998,Podsiadlowski2002,Nelson2003}.
\cite{Lattimer2007} showed that the masses of some NSs are lower
than 1.4 $M_\odot$. Recently, \cite{Meer2007} found that the masses
of NSs in SMC X-1 and Cen X-3 are $1.06^{+0.11}_{-0.10}$ $M_\odot$
and $1.34^{+0.16}_{-0.14}$ $M_\odot$, respectively. However, It is
well known that most of the accurately measured masses of NSs are
near 1.4 $M_\odot$.

In our work, we use the initial masses of NSs via CCSN in
\cite{Hurley2000} and $M_{\rm NS}^{\rm i}=1.4 M_\odot$ in different
simulations, respectively. For NSs via AIC, following
\cite{Hurley2000}, we take $M_{\rm NS}^{\rm i}=1.3 M_\odot$.
Similarly, for NSs via EIC, we also take $M_{\rm NS}^{\rm i}=1.3
M_\odot$.

In addition, nascent NS receives additional velocity  (``kick'') due
to some still unclear process that disrupts spherical symmetry
during the collapse or later Dichotomous nature of kicks which was
suggested quite early by \citet{Katz1975}. Observationally, the kick
is not well constrained due to numerous selection effects.
Currently, high kicks ($\sim100\,\rm km\ s^{-1}$) are associated
with NS originating from CCSN, while low kicks ($\sim10\rm km\
s^{-1}$) with NS born in EIC and AIC \citep{Pfahl2002}. We apply to
core-collapse NS Maxwellian distribution of kick velocity $v_{\rm
k}$
\begin{equation}
P(v_{\rm k})=\sqrt{\frac{2}{\pi}}\frac{v^2_{\rm k}}{\sigma^3_{\rm
k}}e^{-v^2_{\rm k}/2\sigma^2_{\rm k}}.
\end{equation}
Variation of $\sigma_{\rm k}$ between 50 and 200 km s$^{-1}$,
introduces an uncertainty $\lesssim 3$ in the birthrate of low- and
intermediate-mass X-ray binaries \citep{Pfahl2003}. \cite{Zhu2012}
discussed the effects of parameter $\sigma_{\rm k}$ on LMXBs'
populations. Since in this paper we focus on the physical parameters
that mostly affect the masses of NSs,  we do not discuss the effects
of $\sigma_{\rm k}$ on SSs' population. Following \cite{Lu2012}, we
take $\sigma_{\rm k}=190$ km s$^{-1}$ in CCSN, and  $\sigma_{\rm
k}=20$ km s$^{-1}$ in EIC and AIC.

\subsection{Mass of Accreting NS}
In LMXBs, NSs accrete the matter from their companions via Roche
lobe flows or stellar winds. The interaction of a rotating
magnetized NS (single or in a binary system) with surrounding matter
has been studies by many authors \citep[e.
g.,]{Pringle1972,Illarionov1975,Ghosh1978,Lovelace1995,Lovelace1999}.

Using a convenient way of describing NS evolution elaborated by
\cite{Lipunov1992} and  a recent model for quasi-spherical accretion
including subsonic settling proposed by \cite{Shakura2012},
\cite{Lu2012} gave detailed simulations for spin period evolution
and matter accretion of NSs in binaries. In this work, we adopt
their model. \cite{Lu2012} assumed that all matter transferred is
accreted by the NS in an accretor stage. We introduce a parameter
$\beta$ which is the fraction of transferred matter accreted by the
NS, and the rest of the transferred matter is lost from binary
system. The lost matter takes away the specific angular momentum of
the prospective donor. The value of $\beta$ has been usually set to
0.5 \citep{Podsiadlowski1992,Podsiadlowski2002,Nelson2003}. In our
work, we set $\beta=0.1$, 0.5 and 1.0 in different simulations.

\subsection{LMXBs with Nascent SSs}
As noted in \S \ref{sec:introduction}, when the gravitational mass
of a NS reaches to $\sim 1.8 M_\odot$,  the NS can turn into a SS
via the hydrodynamically unstable conversion or the slow conversion.
In the former mode, the binary system may be disrupted
\citep{Ouyed2011b}, and it becomes two isolated stars. One of them
is an isolated SS. However, it is difficult to know the effects of
the hydrodynamically unstable conversion on binary systems.
Therefore, in our work, we consider two extreme cases: (i) In order
to simulate all potential properties of qLMXBs, we assumed all LMXBs
are not affected and survive after the conversion, that is, LMXBs
become qLMXBs when the masses of accreting NSs are larger than $1.8
M_\odot$; (ii) We assumed that all LMXBs are disrupted after the
conversion, that is, there is no qLMXBs. However, we can discuss the
origin of isolated submillisecond pulsars.

In our simulations, if a nascent NS has larger mass than 1.8
$M_\odot$ it is a SS. Therefore, in the paper, qLMXBs include some
LMXBs in which the nascent NSs have larger masses than 1.8
$M_\odot$. \cite{Lai2009} suggested that SSs could have high maximum
masses (See Figure 2 in \cite{Lai2009}) if they are composed of the
Lennard-Jones matter. In our work, we assume that the maximum mass
of SS is 3.0 $M_\odot$.

\section{Results}
We use Monte Carlo method to simulate the initial binaries. For
initial mass function, mass-ratios, and separations of components in
binary systems, we adopt the distributions used by us in
\cite{Lu2006,Lu2008}. We assume that all binaries have initially
circular orbits. After a supernova, new parameters of the orbit are
derived using standard formulae\citep[e. g., ][]{Hurley2002}. It is
well known that theoretical models of the population of LMXBs depend
on badly known input parameters, such as kick velocity and common
envelope treatment \citep[e. g., ][]{Pfahl2003,Zhu2012}. However, in
this pioneering study of qLMXBs we focus on the effects which are
important for the masses of NSs:  the efficiency accreted of
transferred matter, $\beta$ ( $\beta=0.1$, 0.5 and 1.0 ), and the
mass of nascent NS via CCSN (See Figure \ref{fig:nsm} ). We use
$1\times10^8$ binary systems in our Monte-Carlo simulations.

\subsection{Mass Increases of Accreting NSs }
\begin{figure}
\includegraphics[totalheight=3.3in,width=3.3in,angle=-90]{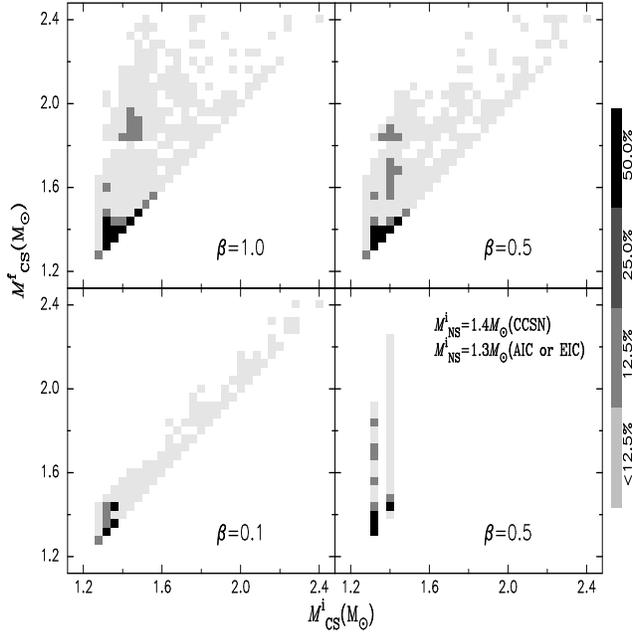}
\caption{Distributions of the final masses of compact stars (NSs
         or SSs) vs. their initial masses in LMXBs for different input parameters.
          Gradations of gray-scale correspond to the number density of systems
            $>$1/2, 1/2 -- 1/4, 1/4 -- 1/8, 1/8 -- 0 of the maximum of
             ${{{\partial^2{N}}\over{\partial {M_{\rm CS}^{\rm i}}{\partial {M_{\rm CS}^{\rm f}}}}}}$
              and blank regions do not contain any stars.}\label{fig:mams}
\end{figure}

Figure \ref{fig:mams} gives the mass increases of accreting NSs in
LMXBs. According the assumptions of SSs formatting from NSs with
masses higher than 1.8 $M_\odot$, 1$\permil$ ($\beta=0.1$)
--- 10\% ($\beta=1.0$) of LMXBs are qLMXBs in our simulations.
This proportion greatly depends on input parameters $\beta$ and the
initial masses of nascent NSs. In the cases of $\beta=1.0$ and
$\beta=0.5$, most of SSs in qLMXBs come from low-mass NSs with 1.4
$M_\odot$, that is, they have accreted $\sim 0.4 M_\odot$ matter. In
the case of $\beta=0.1$, most of SSs in qLMXBs originate from CCSN.

Mass increases of accreting NSs in LMXBs depend on not only input
parameter $\beta$ in this work, but also orbital periods and NSs'
companions. \cite{Zhu2012} showed that most of NSs in LMXBs with WD
donors have low mass-accretion rates ($\sim 10^{-12}\dot{M}_{\odot}
{\rm yr}^{-1}$) and most of LMXBs with WD donors are transient.
Therefore, the masses of accreting NSs in LMXBs with WD donors
hardly reach to 1.8 $M_\odot$. Less than 10\% (in case of
$\beta=0.1$)--- 1$\permil$ (in case of $\beta=0.5, M_{\rm NS}^{\rm
i}=1.4 M_\odot$) of qLMXBs have WD donors in our simulations. Most
of qLMXBs have main sequence donors. From now on, we just discuss
LMXBs and qLMXBs with MS donors.

\subsection{Properties of qLMXBs }
According to our assumption that the conversion of NS matter to SS
matter does not affect binary systems, we can simulate some
observational properties of LMXBs and qLMXBs, and wish to find some
differences between them. Then, taking the case of $\beta=0.5$ as an
example, we discuss some properties of qLMXBs.

\begin{figure}
\includegraphics[totalheight=3.3in,width=3.3in,angle=-90]{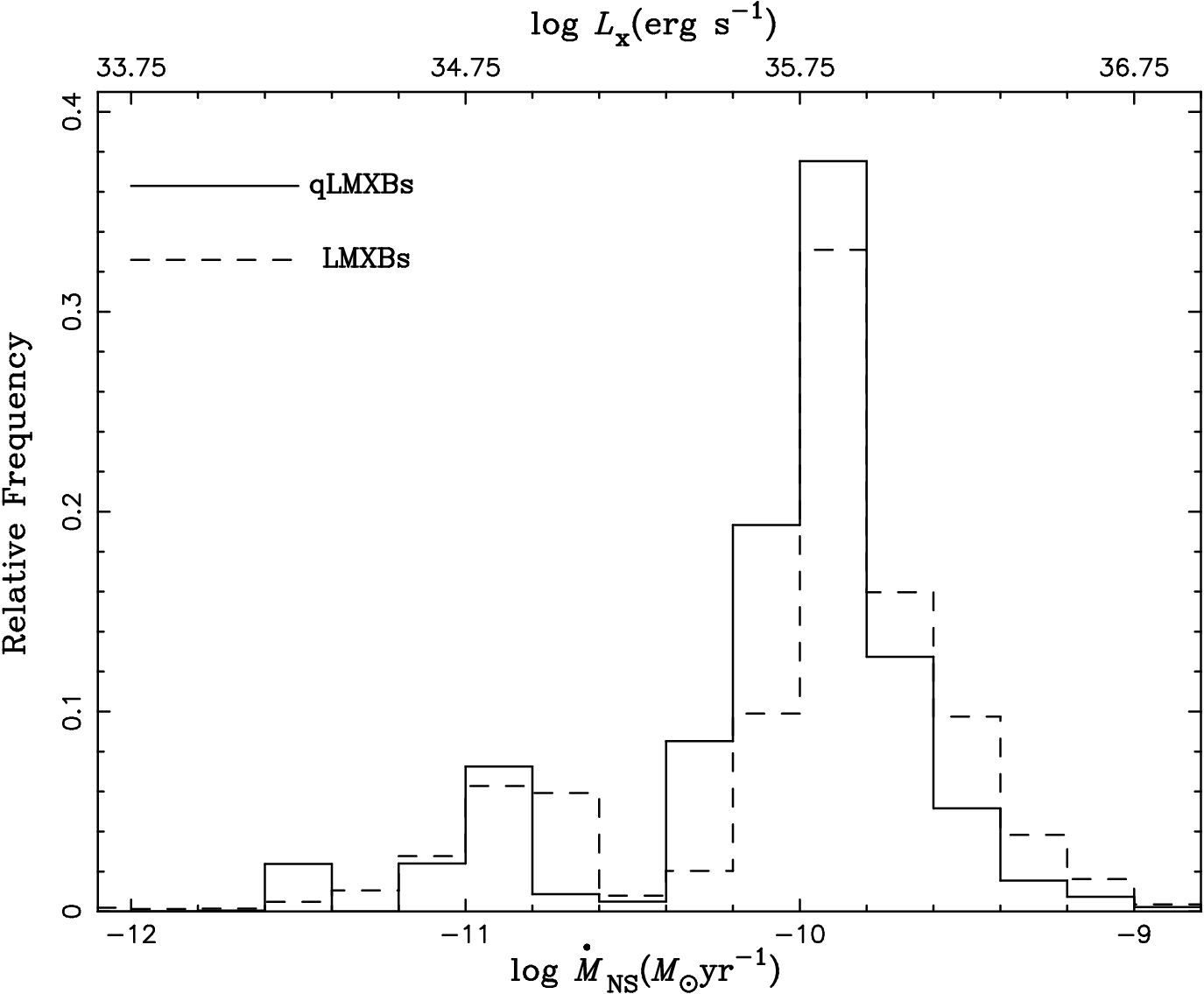}
\caption{Distributions of accretion rates (X-ray luminosities) by
NSs or SSs in LMXBs and qLMXBs, respectively.}\label{fig:ma}
\end{figure}

X-ray luminosities (mass-accretion rates), spin periods and orbital
periods are important parameters of LMXBs. Figure \ref{fig:ma} shows
the accretion rates by NSs or SSs in LMXBs and qLMXBs, and the X-ray
luminosities which are approximated as
\begin{equation}
\begin{array}{l}
L_{\rm x} = \eta \dot{M}_{\rm NS}{\rm c}^2 \simeq
5.7\times10^{35}{\rm erg \
s^{-1}}(\frac{\eta}{0.1})\times(\frac{\dot{M}_{\rm NS}}{{10^{-10}{
M_\odot {\rm yr^{-1}}}}}),
\end{array}\label{eq:lx}
\end{equation}
where $\eta \simeq 0.1$ is the efficiency of converting accreted
mass into X-ray photons. We can find that there is not significant
difference between accretion rates by NSs or SSs in LMXBs and
qLMXBs.

\begin{figure}
\includegraphics[totalheight=3.3in,width=3.3in,angle=-90]{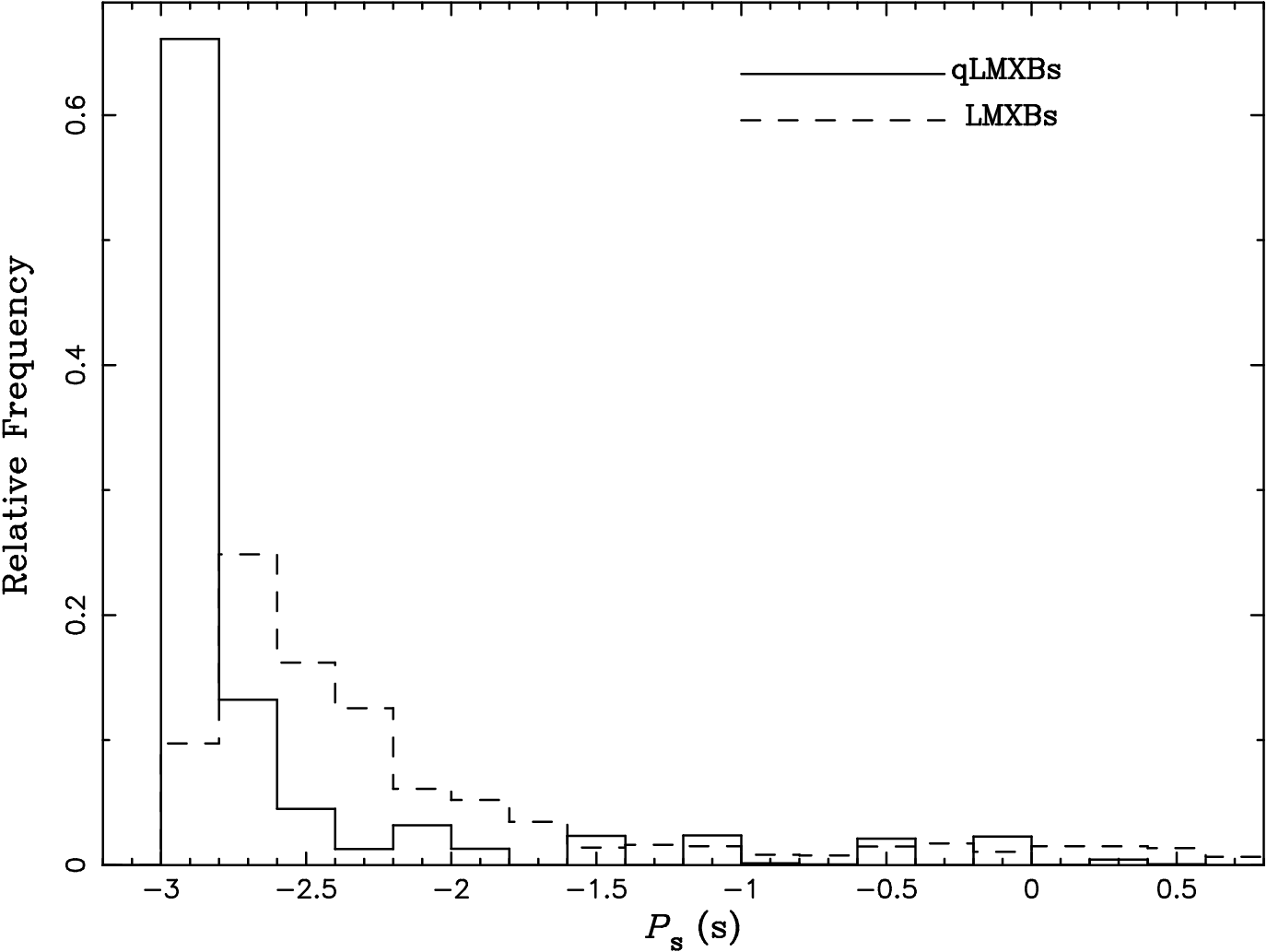}
\caption{Number distributions of the spin periods of NSs or SSs in
LMXBs and qLMXBs, respectively.  The numbers are normalized to
1.}\label{fig:spin}
\end{figure}

The spin periods of the rotating magnetized NSs mainly depend on
their mass-accretion rates \citep{Lu2012}. If accreting SSs are
similar to NSs, we can simulate the spin periods of accreting SSs.
Figure \ref{fig:spin} gives the distribution of the spin periods of
NSs in LMXBs or SSs in qLMXBs. In general, spin periods of SSs in
qLMXBs are shorter than those of NSs in LMXBs.

\begin{figure}
\includegraphics[totalheight=3.3in,width=3.3in,angle=-90]{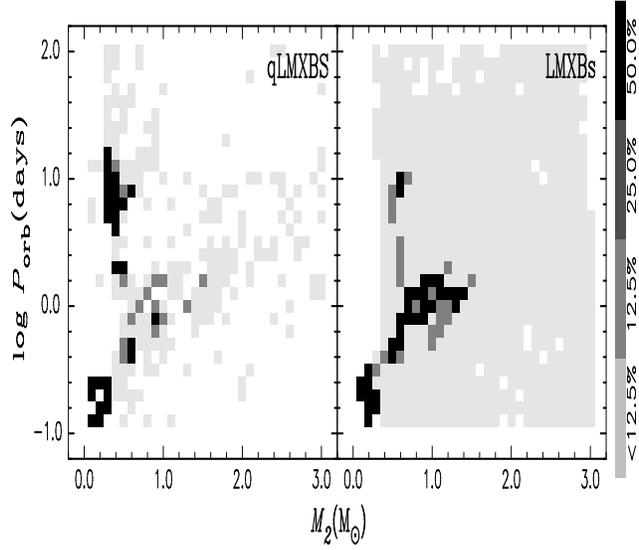}
\caption{Similar to Figure \ref{fig:mams}, but for the distributions
         of the orbital periods $P_{\rm orb}$ of qLMXBs and LMXBs vs. masses of their secondary stars,
         respectively.}\label{fig:mmpp}
\end{figure}

Figure \ref{fig:mmpp} shows the  the distributions of the orbital
periods $P_{\rm orb}$ of qLMXBs and LMXBs vs. masses of their
secondary stars. Similarly, there is not significant difference
between qLMXBs and LMXBs.

As the above descriptions show, it is very difficult in our model to
differ from LMXBs and qLMXBs. The most effective way is to measure
the masses of compact objects in LMXBs if the assumption that NSs
with masses larger than 1.8 $M_\odot$ are SSs is right.

\cite{Jonker2005} suggested that compact object in 2S 0921-630 (It
is a LMXB) has a mass between $\sim$1.9---2.9 $M_\odot$. According
to our assumption, the compact object is a SS. However,
\cite{Steeghs2007} considered that \cite{Jonker2005} overestimated
the rotational broadening and the mass of compact object in 2S
0921-630 is $\sim$1.44 $M_\odot$. The orbital period of 2S 0921-630
is 9.006$\pm$0.007 days, and its X-ray luminosity is $\sim 10^{36}$
erg s$^{-1}$ \citep{Kallman2003}. Results of simulating LMXBs and
qLMXBs in our work both cover these observations. Therefore, we can
not conclude whether compact object in 2S 0921-630 is NS or SS.

\cite{Demorest2010} gave that PSR 1614-2230 has a mass of 1.97
$M_\odot$. And, it is a millisecond radio pulsar (Pulsar spin period
is 3.15 ms) in an 8.7 day orbit, and its companion has a mass of 0.5
$M_\odot$. Although PSR 1614-2230 is not X-ray binary, it may come
from X-ray binary. \cite{Lin2011} suggested that PSR 1614-2230
descended from a LMXB very much like Cyg X-2 ($P_{\rm orb}$ =9.8
days, $M_{\rm NS}=1.7 M_\odot$ and $M_2=0.6 M_\odot$, see
\cite{Casares2010}). AS Figure \ref{fig:mmpp} shows, our results
cover the positions of orbital periods and companion masses of Cyg
X-2 and radio millisecond pulsar binary PSR 1614-2230. Our work
support that PSR 1614-2230 originate from a LMXB. PSR 1614-2230 may
be a SS.
%SAX J1808.4-3658 is an accretion millisecond X-ray pulsar, and it
%spin period is 2.5 ms. The orbital period of the system is 2.014167
%hours. Analysis of the bursts in SAX J1808.4-3658 indicates that it
%is 4 kpc distant and has a peak x-ray luminosity of $6\times10^{36}$
%erg s$^{-1}$ in its bright state, and $<10^{35}$ erg s$^{-1}$ in
%quiescence \citep{Zand1998}.

\subsection{Submillisecond Pulsars }
\cite{Weber2005} suggested that an isolated  submillisecond pulsar
spinning at $\sim$ 0.5 ms could strongly hint the existence of SSs.
If the conversion of NS matter to SS matter is hydrodynamically
unstable and LMXBs are disrupted, the nascent SSs are isolated.
Assuming that one binary with $M_1 \geq 0.8 M_\odot$ is formed per
year in the Galaxy \citep{Yungelson1993, Han1998, Hurley2002}, we
can estimate that the occurrence rate of hydrodynamically unstable
conversion is about 5---70 per Myr.

\begin{figure}
\includegraphics[totalheight=3.3in,width=3.3in,angle=-90]{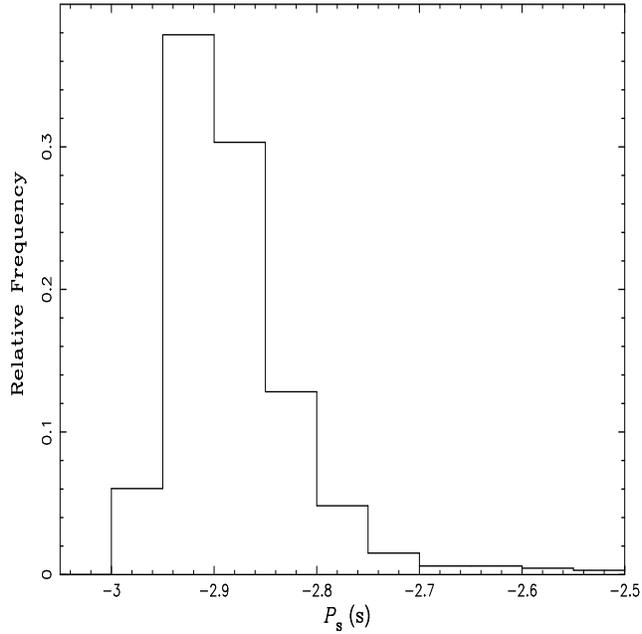}
\caption{Number distribution of the spin periods of NSs at the
beginning of the conversion of NS matter to SS matter.
}\label{fig:qksp}
\end{figure}

As Figure \ref{fig:qksp} shows, the majority of NSs at the beginning
of the conversion have spin periods longer than 1 ms. If the angular
momentum is conserved and no unknown physical mechanic spins up NSs
and nascent SSs during the conversion, the spin periods of nascent
SSs depend on the change of moment of inertia. \cite{Ouyed2011b}
considered that a typical ejected mass of hydrodynamically unstable
conversion is $\sim 10^{-3}M_\odot$. Therefore, the change of moment
of inertia is determined by the difference of radius between NS and
SS. \cite{Ouyed2002} estimated that NS could shrink by as much as
30\%. Then, many nascent SSs have spin periods of $\sim 0.5$ ms in
our simulations. However, this result greatly depends on the
equation of state of NSs and SSs which are poorly known. If NS only
shrinks by as much as 10\% (Private discussion with Xu), it is
difficult for the nascent SSs to spin up to  0.5 ms.

\section{Conclusions}
Employing the population synthesis approach to the evolution of
binaries and using the interacting model of a rotating magnetized NS
with surrounding matter, we investigate the mass change of NSs in
LMXBs and the possibility from NSs converting SSs in LMXBs. Our
results show that about 1\permil --- 10\% of LMXBs can produce SSs.
These SSs may exist in qLMXBs or be isolated, which depends on
physical model of the conversion of NS matter to SS mater.

Our toy model can not conclude whether there are SSs in the Galaxy
and can not give what properties qLMXBs have. In further work, we
need detailed physical model (equation of state about NS and SS and
the conversion of NS matter to SS mater) to improve our work.

\section*{Acknowledgments}
This work was supported by the National Natural Science Foundation
of China under Nos. 11063002 and 11163005, Natural Science
Foundation of Xinjiang under Nos.2009211B01 and 2010211B05,
Foundation of Huoyingdong under No. 121107, Foundation of Ministry
of Education under No. 211198, and Doctor Foundation of Xinjiang
University (BS100106).
%\input refs.tex

%%%%%%%%%%%%%%%%%%%%%%%%%%%%%%%%%%%%%%%%%%%%%%%%%%%%%%%%%%%
%%%%%%%%%%%%%%%%%%%%%%%%%%%%%%%%%%%%%%%%%%%%%%%%%%%%%%%%%%%%
%\bibliography{guoliangads}
\bibliographystyle{apj}
\bibliography{lglmn,lglmn1_pk}
%\begin{thebibliography}{99}
\label{lastpage}
%\end{thebibliography}
\end{document}